\documentclass[prl,twocolumn,preprintnumbers,amsmath,amssymb]{revtex4}
\usepackage{graphicx}
\usepackage{dcolumn}
\usepackage{bm}
\usepackage{subfigure}
\usepackage[dvips]{color}
\allowdisplaybreaks[4]
\newcommand{\beq}{\begin{equation}}
\newcommand{\eeq}{\end{equation}}
\newcommand{\bq}{\begin{equation}}
\newcommand{\eq}{\end{equation}}
\newcommand{\ba}{\begin{array}}
\newcommand{\ea}{\end{array}}
\newcommand{\beqa}{\begin{eqnarray}}
\newcommand{\eeqa}{\end{eqnarray}}
\newcommand{\beqs}{\begin{subequations}}
\newcommand{\eeqs}{\end{subequations}}
\newcommand{\infinity}{\infty}

\def\nn{\nonumber}

\def\geqq{\geqslant}
\def\cut{\Lambda}
\def\k{\kappa}
\def\kk{\kappa^2}

\def\nn{\nonumber}
\def\f{\frac}
\def\dis{\displaystyle}
\def\[{\left[}
\def\]{\right]}
\def\({\left(}
\def\){\right)}
\def\dif{\partial}
\def\hf{\frac{1}{2}}

\def\phih{\widehat{\varphi}}
\def\phihh{\hat{\phi}}
\def\phibar{\bar{\phi}}

\def\varphib{\bar{\varphi}}
\def\II{\mathcal{I}}

\evensidemargin=\oddsidemargin

\def\la{\lambda}
\def\al{\alpha}
\def\be{\beta}

\def\si{\sigma}
\def\sigmab{\bar{\sigma}}
\def\sigmah{\hat{\sigma}}

\def\at{\tilde{a}_0^{}}
\def\bt{\tilde{b}_0^{}}

\def\pslash{\not{\hbox{\kern-4pt $p$}}}
\def\qslash{\not{\hbox{\kern-4pt $q$}}}
\def\lv{\not{\hbox{\kern-4pt $L$}}}
\def\lsim{\mathrel{\raise.3ex\hbox{$<$\kern-.75em\lower1ex\hbox{$\sim$}}}}
\def\gsim{\mathrel{\raise.3ex\hbox{$>$\kern-.75em\lower1ex\hbox{$\sim$}}}}
\def\ifmath#1{\relax\ifmmode #1\else $#1$\fi}

\def\End{\end{document}}

\begin{document}

 \title{\large Gauge-Invariant Quantum Gravity Corrections to Gauge Couplings\\
               via Vilkovisky-DeWitt Method and Gravity Assisted Gauge Unification
       }

 \author{\sc  Hong-Jian He$^{1,2}$, Xu-Feng Wang$^{1,3}$, Zhong-Zhi Xianyu$^{1}$}

 \affiliation{
       $^1$\,Center for High Energy Physics and Institute of Modern Physics,
       Tsinghua University, Beijing 100084, China\\
       $^2$\,Kavli Institute for Theoretical Physics China,
       Chinese Academy of Sciences, Beijing 100190, China\\
       $^3$\,Institute for the Physics and Mathematics of the Universe, University of Tokyo,
       Chiba 277-8568, Japan
       }

 \begin{abstract}
 Gravity is the weakest force in nature,
 and the gravitational interactions with all standard model (SM)
 particles can be well described by perturbative expansions of the Einstein-Hilbert action
 as an effective theory, all the way up to energies below the fundamental Planck scale.
 We use Vilkovisky-DeWitt method to derive the first gauge-invariant nonzero gravitational
 power-law corrections to the running gauge couplings,
 which make both Abel and non-Abel gauge interactions asymptotically free.
 We further demonstrate that the graviton-induced universal power-law runnings always assist
 the three SM gauge forces to reach unification at the Planck scale,
 irrespective of the detail of logarithmic corrections.
 We also compute the power-law corrections to the SM Higgs sector
 and derive modified triviality bound on the Higgs boson mass.
 \\[2mm]
  PACS: 12.10.Kt, 04.60.-m, 11.10.Hi.  \hfill
  [\,{\tt arXiv:1008.1839}\,]
 \end{abstract}

 \maketitle

 \noindent
 \section{1. Introduction}

 Although gravity, as the weakest force in nature, is more perturbabtive than
 the other three fundamental forces all the way up to energies below the Planck scale,
 it was found to be non-renormalizable in the conventional sense\,\cite{tHooft-Veltman}.
 But this does not prevent the enormous range of
 successful physical and astrophysical applications of the Einstein general relativity
 of gravitation. In fact, all nature's four fundamental forces can be well described by the
 modern formulation of {\it effective field theories}\,\cite{Weinberg-EFT}, with no exception to
 gravitation\,\cite{EFT-GR}.  The leading terms in the Einstein-Hilbert action,
 \beqa
 \label{eq:S-EH}
 S_{\rm EH} = \int\!\! d^4x \sqrt{-g}\,\kappa^{-2}(R - 2\Lambda_0)\,,\,
 \eeqa
 are just the least suppressed operators in the effective theory of general relativity
 under perturbative low energy expansion, where
 $\,\kappa^2 \equiv {16\pi G}\equiv 16\pi /M_P^2\,$ is fixed
 by the Newton constant $G$ (or Planck mass $M_P\simeq 1.2\times 10^{19}$\,GeV)
 and $\Lambda_0$ denotes the cosmological constant.

 All standard model (SM) particles must join gravitational interaction with their couplings
 controlled by the universal Newton constant $G$. It is thus important to understand,
 under the effective theory formulation, how gravity corrects the SM observables,
 in connection to the other three gauge forces in nature.
 Robinson and Wilczek\,\cite{Wilczek} initiated a very interesting study of gravitational
 corrections to running gauge couplings,  but it was then realized that their
 calculation by using conventional background field method (BFM)\,\cite{BFM}
 is generally gauge-dependent and the net result vanishes\,\cite{gauge-dep,gauge-dep2}.

 However, it is important to note that Vilkovisky and DeWitt\,\cite{Vilkovisky-DeWitt}
 proposed a new approach over the conventional BFM,
 especially powerful for analyses involving gravitation,
 which is guaranteed to be gauge-invariant, independent of the choices of both
 gauge-condition and gauge-parameter\,\cite{Toms-book,VDold}.
 The Vilkovisky-DeWitt method was recently applied by Toms to study logarithmic
 corrections of graviton to the running coupling of QED with a nonzero
 cosmological constant\,\cite{Toms-CC} under dimensional regularization,
 and to scalar mass\,\cite{Toms-phi}.

 The purpose of the present work is to apply the fully gauge-invariant
 Vilkovisky-DeWitt method\,\cite{Vilkovisky-DeWitt} for studying the gravitational corrections to
 the power-law running of Abel and non-Abel gauge couplings.
 We derive {\it the first gauge-invariant nonzero power-law correction,
 which is asymptotically free,} in support of what Robinson and Wilczek hoped.
 We also extend this approach for studying power-law corrections to the SM Higgs sector
 and derive modified triviality bound on the Higgs boson mass\,\cite{paper2},
 as will be summarized in the last part of this paper.
 The power-law running originates from the quadratical divergences
 associated with graviton loop with overall couplings proportional to $\kappa^2$.
 The gravitational coupling $\kappa^2$ has
 negative mass-dimension equal $-2$\,,\, so the graviton induced loop contributions can generate
 generic dimensionless power-law corrections to a given gauge coupling $\,g_i^{}\,$,\,
 of the form $\,g_i^{}\kappa^2\Lambda^2\,$,
 where $\Lambda$ is the ultraviolet (UV) momentum cutoff. After renormalization one can deduce
 the generic form of one-loop Callan-Symanzik $\beta$ function by general dimensional analysis,
 \beqa
 \label{eq:beta}
 \beta (g_i^{},\mu )
 ~=\, -\f{b_{0i}}{(4\pi)^2}g_i^3 +\f{a_0^{}}{(4\pi)^2}(\kappa^2 \mu^2)g_i^{} \,,
 \eeqa
 where $\mu$ is the renormalization scale and the coefficient $a_0^{}$ has to be determined by explicit,
 gauge-invariant computation of graviton radiative corrections.
 There is no reason {\it a priori} to expect $\,a_0^{}\,$
 be exactly zero (as stressed by Robinson and Wilczek).
 Our key point here is to extract the physical
 power-law corrections via a fully gauge-invariant method
 \`{a} la Vilkovisky-DeWitt\,\cite{Vilkovisky-DeWitt}.
 The physical meaning of the quadratical divergences in nonrenormalizable theories was clarified
 in depth by Veltman\,\cite{Veltman} and he advocated to use dimensional reduction (DRED)
 method \cite{DRED} (rather than dimensional regularization (DREG))
 to consistently regularize quadratical divergences
 as $\,d=2\,$ poles for the Higgs mass corrections in the SM.
 Then, Einhorn and Jones made further insight\,\cite{Einhorn} that any regularization procedure
 which preserves the right number of spin degrees of freedom for each field should give
 the correct results of quadratical divergence. This includes DRED but excludes DREG, as
 DREG miscounts the physical spin degrees
 of freedom for dealing with quadratical divergences\,\cite{Veltman,Einhorn}.
 For quadratically divergent integrals, except to know that a consistent regularization such as
 DRED exists for them, there is no need to explicitly work out these integrals
 until after we finish computing and summing up all their coefficients via gauge-invariant formulation.
 Then we can identify the remaining single divergent integral and re-regularize it at $\,d=4\,$
 by placing a common physical momentum cutoff; the renormalization will be carried out to extract
 the power-law corrections.
 (This procedure was applied to extract the gauge-invariant quadratical divergence
 in the Higgs boson mass and was shown to be regularization-independent\,\cite{Einhorn}.)
 We have explicitly used DRED method for our analysis (\`{a} la Veltman\,\cite{Veltman})
 and checked all possible consistencies.

 \section{2. Gauge-Invariant Vilkovisky-DeWitt Effective Action}

 The Vilkovisky-DeWitt approach\,\cite{Vilkovisky-DeWitt} modifies
 the conventional BFM in order to build a manifestly gauge invariant effective action.
 It is noted that a gauge transformation corresponds to a field-reparametrization in field space
 $\,\varphi_{i}^{}\to\varphi^{\prime}_{i}\,$
 (where we use the condensed notation of DeWitt\,\cite{DeWitt}, with
 the subscript $i$ denoting all internal and Lorentz indices besides the spacetime coordinates).
 A variation of the gauge-fixing condition for a gauge theory is equivalent
 to a change in the external source term.
 This means that although the classical action $S[\varphi]$ is invariant under field-reparametrization,
 the effective action $\,\Gamma[\bar\varphi]\,$ in the conventional BFM is not a scalar.
 The definition of conventional $\,\Gamma[\bar\varphi]\,$,
 \begin{equation}
 \label{eq:old-Gamma}
 e^{i\Gamma[\bar{\varphi}]}
 =\int\!\! d\varphi\mu[\varphi]
  \exp i\!\left[S[\varphi]+(\bar{\varphi}^i\!-\!\varphi^i)
  \frac{\delta\Gamma[\bar{\varphi}]}{\delta\bar{\varphi}^i}\right]\!,~~~
 \end{equation}
 includes the external source $\,\delta\Gamma[\varphib ]/\delta\varphib^i =-J_i\,$,\,
 where $\,\bar{\varphi}^{i}\,$ denotes the background of $\,\varphi^{i}$,\, and
 $\,d\varphi\,\mu[\varphi]$\, is the measure of functional integral. For a gauge theory,
 $\,\mu[\varphi]\,$ contains gauge-fixing condition
 and the corresponding DeWitt-Faddeev-Popov determinant.
 From the geometrical viewpoint with the field configuration
 space as a manifold\,\cite{Vilkovisky-DeWitt},
 it is clear that the difference of two distinct points
 $\,\bar\varphi^i-\varphi^i\,$ in (\ref{eq:old-Gamma}) is
 not field-reparametrization covariant, and thus $\Gamma[\bar{\varphi}]$
 may depend on the choice of gauge condition.
 \begin{figure}
 \begin{center}
\includegraphics[width=0.23\textwidth]{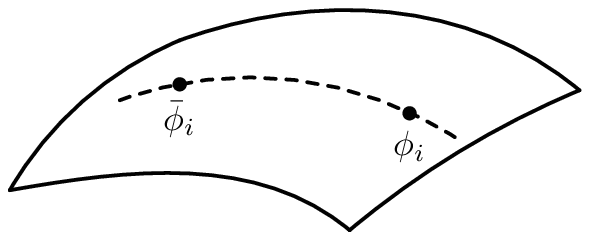}
\includegraphics[width=0.23\textwidth]{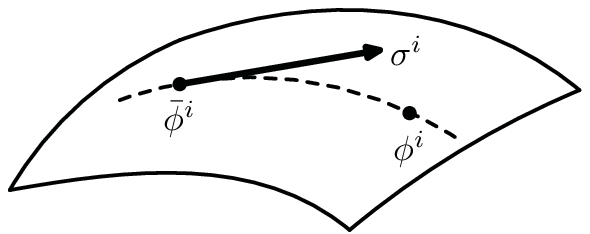}
   \caption{Coordinates difference in field space as a manifold.}
   \label{field-space}
 \vspace*{-4mm}
 \end{center}
 \end{figure}

 The way out of this trouble is to replace the coordinate difference
 $\,\bar\varphi^i-\varphi^i\,$ by a covariant vector
 $\,\sigma^i[\varphi,\bar\varphi]\,$ as illustrated in Fig.\,\ref{field-space},
 and introduce a connection $\Gamma^i_{jk}$ in field space,
 we can expand $\sigma^i$ as,
 \begin{equation}
 \sigma^i[\varphib,\varphi] =
 \varphib^i-\varphi^i - \f{1}{2}\Gamma^i_{jk}
 (\varphib - \varphi )^j (\varphib - \varphi )^k + \cdots .
 \end{equation}
 Thus, the Vilkovisky-DeWitt effective action $\,\Gamma_{\text G}$\,
 can be constructed as a scalar in field space \cite{Vilkovisky-DeWitt},
 \begin{equation}
 \label{eq:Gamma-G-sigma}
 e^{i\Gamma_{\text G}[\bar{\varphi}]}
 = \!\!\int\!\! d\varphi\mu[\varphi]
 \exp i\!\left[S[\varphi]+C^{-1i}_j[\bar{\varphi}]
 \Gamma_{\text{G},i}[\bar{\varphi}]\sigma^j[\bar{\varphi},\varphi]\right],
 \end{equation}
 where
 $\,\Gamma_{\text{G},i}\equiv \delta\Gamma_G/\delta\bar{\varphi}\,$ and
 a subscript ``comma" will be always used to denote the functional derivative.
 The coefficient $C_j^{-1i}$ can also be expanded perturbatively,
 \begin{equation}
 \!
 C_j^{-1i}[\bar\varphi,\varphi] \,=\, \delta^i_j+\frac{1}{3}
 R^i_{mnj}[\bar\varphi]\sigma^m[\bar\varphi,\varphi]
 \sigma^n[\bar\varphi,\varphi]+\cdots ,~~
 \end{equation}
 where $R^i_{mnj}$ is the curvature tensor associated with connection
 $\,\Gamma^i_{jk}$\,.\,  Since the $R^i_{mnj}$-term (or other higher order terms in the
 expansion) already contains two covariant vectors $\si^m$ and $\si^n$ (or more),
 it could contribute to the effective
 action (\ref{eq:Gamma-G-sigma}) a term which is at least cubic in $\,\si^i$\,,\,
 and thus will be irrelevant to the one-loop effective action.

 Under perturbation expansion, we can write down the
 one-loop Vilkovisky-DeWitt effective action, which
 is also a scalar under reparametrization,
 \begin{equation}
 \Gamma_G[\bar{\varphi}]=S[\bar{\varphi}]-i\ln\mu[\bar\varphi]
 +\frac{i}{2}\textrm{Tr}\ln\nabla_m\nabla_nS \,,~~~
 \end{equation}
 where $\,\nabla_{m}\,$ is the covariant derivative associated
 with connection $\,\Gamma^i_{mn}\,$,
 \begin{equation}
 \frac{\delta^2S}{\delta\varphi^m\delta\varphi^n}
 ~\to~
 \nabla_m\nabla_nS=\frac{\delta^2S}{\delta\varphi^m\delta\varphi^n}-\Gamma^i_{mn}(\varphi)
 \frac{\delta S}{\delta\varphi^i}\,.
 \end{equation}

 For a gauge theory, consider the infinitesimal gauge transformation,
 \begin{equation}
 \delta\varphi^{i} ~=~ K^{i}_{\alpha}[\varphi]\epsilon^{\alpha}\,,
 \end{equation}
 with $K^{i}_{\alpha}[\varphi]$ being the generators of gauge transformation
 and $\epsilon^{\alpha}$ the infinitesimal gauge-group parameters.
 Thus, for quantization the gauge-fixing condition $\,\chi_{\alpha}^{}(\varphi)\,$
 and the DeWitt-Faddeev-Popov ghost term $\,Q_{\alpha\beta}[\bar{\varphi}]=\frac{\delta\chi_{\alpha}^{}}{\delta\epsilon^{\beta}}\,$
 should be introduced.
 So, the Vilkovisky-DeWitt effective action is given by, up to one-loop order,
 \begin{eqnarray}
 \label{1loopgauge}
 \Gamma[\bar{\varphi}]
 &=& S[\bar{\varphi}]-\textrm{Tr}\ln Q_{\alpha\beta}[\bar{\varphi}]
 \nn\\
 & &
 +\frac{i}{2}\textrm{Tr}\ln\!\left(\!\nabla_m\nabla_n S
 +\frac{1}{2\xi}(\chi_\alpha^{}\chi^\alpha)_{,\hat\varphi,\hat\varphi}^{}
 [\bar{\varphi}]\!\right) \!,~~~~~
 \end{eqnarray}
 which is proven to be invariant under the change of gauge condition
 $\,\chi_{\alpha}^{}\,$ and gauge-fixing parameter $\,\xi$\,
 \cite{Vilkovisky-DeWitt,Toms-book}. We have defined the background field $\,\bar{\varphi}^{i}\,$
 and fluctuating field $\,\phih^{i}\,$ via  $\,\varphi^{i}=\bar{\varphi}^{i}+\phih^{i}\,$.
 For calculations in a specific gauge theory, the connection $\,\Gamma^{i}_{jk}\,$
 is very complicated and non-local. But it can be shown that,
 when the Landau-DeWitt gauge condition
 \begin{equation}
 \label{eq:Landau-DeWitt-GC}
  \chi_{\alpha}[\bar{\varphi},\phih]
  ~=~ K_{\alpha i}[\bar{\varphi}]\phih^{i} ~=~ 0
 \end{equation}
 is chosen, the relevant parts of connection coefficients are simply given
 by the Christoffel symbol associated with a metric $G_{ij}$ in
 field space\,\cite{Fradkin-Tseytlin},
 \begin{equation}
 \label{eq:connection-def}
 \Gamma^i_{jk} ~=~ \frac{1}{2}G^{i\ell}(G_{\ell j,k} + G_{\ell k,j} - G_{jk,\ell}) \,.
 \end{equation}

 In summary, Vilkovisky-DeWitt effective action provides a fully gauge-invariant
 description of the off-shell gauge field theories, which is guaranteed to be independent
 of choices of both gauge-condition and gauge-parameter.
 In the following, we will apply this method to analyze the quantum gravity coupled
 to the Abel and non-Abel gauge theories, as well as the SM Higgs sector.

 \section{3. Gravitational Corrections to Abel and Non-Abel $\beta$ Functions}

 We start from the classical action of Einstein-Maxwell theory, which consists of the
 Einstein-Hilbert action (\ref{eq:S-EH}) and
 \begin{equation}
 \label{eq:S-EM}
 S_{\rm EM} ~=\,
 -\frac{1}{4}\int\!\!d^4x\sqrt{-g}\,g^{\mu\alpha}g^{\nu\beta}F_{\mu\nu}F_{\alpha\beta} \,.
 \end{equation}
 Since the Vilkovisky-DeWitt method does not require on-shell background,
 we expand the metric $g_{\mu\nu}$ around the Minkowski background $\,\eta_{\mu\nu}$\,,
 \begin{equation}
 g_{\mu\nu}^{} ~=~ \eta_{\mu\nu}^{} + \kappa h_{\mu\nu}^{} \,.
 \end{equation}
 We further split the gauge field $A_\mu$ as
 \begin{equation}
 A_{\mu}=\bar{A}_{\mu}+a_{\mu}^{} \,,
 \end{equation}
 with $\,\bar{A}_{\mu}\,$ the background field and $a_{\mu}^{}$ the quantum fluctuating field.
 As shown in Eq.\,(\ref{eq:connection-def}), the
 connection $\,\Gamma^i_{jk}\,$ is determined from the metric $G_{ij}$ defined
 in the field manifold. There is a natural choice\,\cite{Vilkovisky-DeWitt}
 of the field-space metric $G_{ij}$ with its nonzero components given by
 \beqa\label{gravity-metric}
 \hspace*{-4mm} &&
 G_{g_{\mu\nu}^{}(x)g_{\al\be}^{}(y)}
 =\frac{\sqrt{-g}}{2\kappa^2}(g^{\mu\al}g^{\nu\be}
  \!+\! g^{\mu\be}g^{\nu\al} \!-\! g^{\mu\nu}g^{\al\be})\delta(x\!-\!y),
 \nn\\[2mm]
 \hspace*{-4mm} &&
 G_{A_\mu(x)A_\nu(y)}=\sqrt{-g}\,g^{\mu\nu}\delta(x\!-\!y)\,.
 \eeqa
 Thus, the relevant part of connection can be derived from Eq.\,(\ref{eq:connection-def})
 under the Landau-DeWitt gauge condition (\ref{eq:Landau-DeWitt-GC}),
 %
 \begin{align}
 \label{connection-qed}
 \hspace*{-3mm}
  \Gamma^{g_{\rho\sigma}^{}(x)}_{g_{\mu\nu}^{}(y)g_{\alpha\beta}^{}(z)}
 =\,&\, \mbox{$\frac{1}{4}$}g^{\mu\nu}\delta_\rho^{(\alpha}\delta^{\beta)}_\sigma
  +\mbox{$\frac{1}{4}$}g^{\alpha\beta}\delta_\rho^{(\mu}\delta^{\nu)}_\sigma
  \!-\delta_{(\rho}^{(\alpha}g^{\beta)(\mu}\delta^{\nu)}_{\sigma)}
 \nonumber\\
 \hspace*{-3mm}
 &
 +\mbox{$\frac{1}{4(d-2)}$} g_{\rho\sigma}
  (2g^{\mu(\alpha}g^{\beta)\nu}\!\!-g^{\mu\nu}g^{\alpha\beta})
  \nonumber\\
 &\times\delta(x\!-\!y)\delta(x\!-\!z) ,
 \nn
 \\[1.5mm]
 \hspace*{-3mm}
 \Gamma^{g_{\mu\nu}^{}(x)}_{A_\alpha(y)A_\beta(z)}
 =\,& \mbox{$\frac{1}{4}$}\kappa^2\delta^\alpha_{(\mu}\delta_{\nu)}^\beta
  \delta(x\!-\!y)\delta(x\!-\!z) ,
 \\[1.5mm]
 \hspace*{-3mm}
 \Gamma^{A_\mu(x)}_{A_\nu(y)g_{\alpha\beta}^{}(z)}
 =\,& \mbox{$\frac{1}{4}$}(\delta^\nu_\mu g^{\alpha\beta}
  \!-\delta^\alpha_\mu g^{\nu\beta}\!-\delta^\beta_\mu g^{\nu\alpha})
 \delta(x\!-\!y)\delta(x\!-\!z)  ,
 \nn
 \end{align}
 %
 where we have introduced the symmetrization notation,
 $\,\delta_\rho^{(\alpha}\delta^{\beta)}_\sigma
  =\f{1}{2}\(\delta_\rho^{\alpha}\delta^{\beta}_\sigma +
             \delta_\rho^{\beta}\delta^{\alpha}_\sigma\) $\,,\,
 and so on.
 The Landau-DeWitt gauge condition should be determined by the gauge transformations,
 \begin{subequations}
 \beqa
 \delta g_{\mu\nu}^{}
 &\!=\!& -g_{\mu\alpha}^{}\partial_\nu\epsilon^\alpha - g_{\nu\alpha}^{}
 \partial_\mu\epsilon^\alpha - \epsilon^\alpha\partial_\alpha g_{\mu\nu}^{}
 \,,~~~~~~~~
 \\[1mm]
 \delta A_\mu  &\!=\!&
 -\partial_\mu\epsilon-A_\nu\partial_\mu\epsilon^\nu-\epsilon^\nu\partial_\nu A_\mu\,,~~
 \eeqa
 \end{subequations}
 with $\,\epsilon^\alpha\,$ being the infinitesimal parameter of gravitational gauge transformation
 and $\epsilon$ the infinitesimal parameter of $U(1)$ gauge transformation.
 Then, the gauge-fixing functions for photon and graviton fields are
 \begin{subequations}
 \label{gauge-condition}
 \beqa
 \label{gauge-condition-a}
 \chi &\!=\!& \partial^\mu a_\mu \,,
 \\[1mm]
 \label{gauge-condition-b}
 \chi_\mu^{} &\!=\!& \(\partial^\lambda h_{\mu\lambda}
  -\frac{1}{2}\partial_\mu h\) + \f{\k}{2}a^\lambda\bar{F}_{\mu\lambda}
 \,,~~~~~~~~
 \eeqa
 \end{subequations}
 where $\,\bar{F}_{\mu\nu}=\partial_{\mu}\bar{A}_{\nu}-\partial_{\nu}\bar{A}_{\mu}\,$
 and an overall factor $\,-2/\k\,$ is factorized out in (\ref{gauge-condition-b})
 for the convenience of normalization.
 So the Lagrangian contains the following gauge-fixing terms,
 \begin{equation}
 \mathcal{L}_{\rm gf} ~=~ \frac{1}{2\zeta}\chi_\mu^{}\chi^\mu - \frac{1}{2\xi}\chi^2 \,,
 \end{equation}
 where $\,\xi\,$ (\,$\zeta$\,)\, is gauge-fixing parameter for photon (graviton) field, and will be
 set to zero at the end of calculation, as required by imposing
 the Landau-DeWitt gauge condition.

 Then, we consider the connection-induced terms in the Lagrangian,
 \begin{eqnarray}
 \label{eq:L-con}
 \mathcal{L}_{\rm con}
 &\!\!=\!\!&
 -\frac{1}{2}\Gamma^{g_{\mu\nu}^{}}_{A_\alpha A_\beta}S_{,g_{\mu\nu}^{}}
 a_\alpha^{} a_\beta^{}
 -\f{1}{2}\kappa^2\Gamma^{g_{\rho\sigma}^{}}_{g_{\mu\nu}^{}g_{\alpha\beta}^{}}
 S_{,g_{\rho\sigma}^{}}h_{\mu\nu}h_{\alpha\beta}
 \nonumber\\
 &\!\!\!\!&
 -\kappa\Gamma^{A_\mu}_{A_\nu g_{\alpha\beta}}S_{,A_\mu}a_\nu^{} h_{\alpha\beta} \,,
 \end{eqnarray}
 where
 \beqs
 \beqa
 S_{,g_{\mu\nu}}|_{\bar{\varphi}}
 &\!=\!& -\frac{\Lambda_0}{\kappa^2}\eta^{\mu\nu}\!
  -\frac{1}{8}\eta^{\mu\nu}\bar{F}_{\alpha\beta}\bar{F}^{\alpha\beta}\!
  +\frac{1}{2}\bar{F}^\mu_\lambda \bar{F}^{\nu\lambda} \,,~~~~~~~~~
  \\[1mm]
 S_{,A_\mu}|_{\bar{\varphi}} &\!=\!& \partial_\nu\bar{F}^{\nu\mu} \,.
 \eeqa
 \eeqs
 The ghost part of the Lagrangian is given by
 \begin{equation}
 \mathcal{L}_{\rm gh}
 ~=~ \bar{\eta}^\lambda\delta\chi_\lambda^{} + \bar{\eta}\delta\chi \,,~~~~~~
 \end{equation}
 where $\delta\chi$ and $\delta\chi_\lambda^{}$
 are the changes under gauge transformations with
 $\,\epsilon=\eta\,$ and $\,\epsilon^\lambda=\eta^\lambda\,$,\,
 and $\,(\eta,\,\eta^\lambda)\,$ the anti-commuting ghost fields.

\begin{figure}[h]
  \begin{center}
  \includegraphics[width=0.47\textwidth]{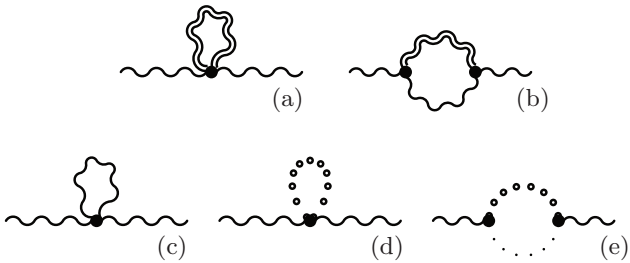}
    \vspace*{-2mm}
    \caption{Graviton induced radiative corrections to photon self-energy, where
         wavy external (internal) lines stand for background (fluctuating) photon fields,
         double-lines for gravitons, dotted lines for photon-ghosts, and
         circle-dotted line for graviton ghosts.}
  \label{photon-self-energy}
  \end{center}
  \end{figure}

 With these we can sum up all required terms for computing the effective action
 (\ref{1loopgauge}),
 \begin{equation}
 \label{action}
 S_Q ~=~ S_{\rm EM} + \int\!\!d^{4}x\,
       (\mathcal{L}_{\rm con}+\mathcal{L}_{\rm gf}+\mathcal{L}_{\rm gh}) \,.
 \end{equation}
 The one-loop effective action will be deduced from
 $\,\f{i}{2}\textrm{Tr}\ln (S_Q)_{,i,j}\,$,\, according to Eq.\,(\ref{1loopgauge}).
 This has clear diagrammatic interpretation.
 For our purpose we are interested in all the bilinear terms of background photon field, which
 correspond to the one-loop self-energy diagrams listed in Fig.\,\ref{photon-self-energy}.

 We note that among all five diagrams in Fig.\,\ref{photon-self-energy}(a)-(e), the first two
 also exist in the conventional BFM or diagrammatic calculation (though the present couplings
 differ from the conventional ones), but the last three arise solely
 from the {\it connection-induced contributions} in the Vilkovisky-DeWitt formulation.
 We systematically compute these diagrams using the Feynman rules from Eq.\,(\ref{action}),
 and extract only the quadratically divergent parts of loop integrals,
 \begin{subequations}
 \label{eq:Ph-SE}
 \beqa
 \label{eq:Ph-SE-a}
 \mathrm{(a)} &=& (1+2\zeta)\kk
 (p^{2}\eta_{\mu\nu}^{}\!-p_{\mu}^{}p_{\nu}^{})\;{\cal I}_{2}\,,
 \\[1mm]
 \label{eq:Ph-SE-b}
 \mathrm{(b)} &=& \(\! -\frac{1\!+\!\xi}{4\zeta}-2-\zeta\)\kk
 (p^{2}\eta_{\mu\nu}^{}\! -p_{\mu}^{}p_{\nu}^{})\;{\cal I}_{2}\,,~~~~~~~~
 \\[1mm]
 \label{eq:Ph-SE-c}
 \mathrm{(c)} &=& \(\frac{1\!+\!\xi}{4\zeta}+\frac{1\!-\!\xi}{8}\)\kk
 (p^{2}\eta_{\mu\nu}^{}\! -p_{\mu}^{}p_{\nu}^{})\;{\cal I}_{2}\,,
 \\[1mm]
 \label{eq:Ph-SE-d}
 \mathrm{(d)} &=& -\kk (p^{2}\eta_{\mu\nu}^{}\! -p_{\mu}^{}p_{\nu}^{})\;{\cal I}_{2}\,,
 \\
 \label{eq:Ph-SE-e}
 \mathrm{(e)} &=& 0\,,
 \eeqa
 \end{subequations}
 where the integral
 \begin{equation}
 \label{eq:I2}
 {\cal I}_{2} ~\equiv~ \int\!\frac{d^{d}k}{(2\pi)^{d}}\frac{1}{k^{2}}
 \end{equation}
 is quadratically divergent for $\,d=4\,$ by power-counting and gets regularized for
 $\,d<2\,$ via DRED with a singular pole at $\,d=2\,$ \cite{Veltman},
 though we need not to explicitly work it out so far\,\cite{Einhorn}.
 Since the external gauge fields in Fig.\,\ref{photon-self-energy} carry Lorentz indices,
 we will encounter some terms containing $\,k^\mu k^\nu\,$ in the integral of
 loop-momentum $\,k$\,,\, where according to the DRED we symmetrize
 $\,k^\mu k^\nu\to \eta^{\mu\nu}k^2/d\,$ with the metric
 $\eta^{\mu\nu}$ defined at $d=4$ and $\,k^2/d\,$ at $\,d\to 2\,$
 for identifying the $1/(d-2)$ poles.
 The physical picture of the DRED is clear\,\cite{DRED,Veltman,Einhorn}:
 the Lorentz indices of loop-momenta in the numerator of the integral should be separated
 into those of the metric $\eta^{\mu\nu}$ or external momenta at $\,d=4\,$ to preserve the
 right spin degrees of freedom for each field, and then the remaining scalar integral
 over the loop-momenta is regularized at $d$ dimension (which can always be reduced to
 Eq.\,(\ref{eq:I2}) for quadratical divergence).
 Summing up all the self-energy contributions (\ref{eq:Ph-SE-a})-(\ref{eq:Ph-SE-e}),
 we find that all $\,\frac{1}{\zeta}\,$ poles explicitly cancel, which is a consistency check
 for Vilkovisky-DeWitt method,  and we deduce the net result in the Landau-DeWitt gauge,
 \begin{equation}
 \label{eq:SE-sum}
 \mathrm{(a)}\!+\!\mathrm{(b)}\!+\!\mathrm{(c)}\!+\!\mathrm{(d)}\!+\!\mathrm{(e)}
 \,=\,-\frac{15}{8}\kk(p^{2}\eta_{\mu\nu}^{}\! -p_{\mu}^{}p_{\nu}^{})\;{\cal I}_{2}\,.~~~~
 \end{equation}
 As guaranteed by the Vilkovisky-DeWitt method\,\cite{Vilkovisky-DeWitt,Toms-book},
 this is a fully gauge-invariant result.
 Noting that the singular pole of the integral (\ref{eq:I2}) at
 $\,d=2\,$ just corresponds to the quadratical divergence of the same integral at $\,d=4\,$
 \cite{Veltman}, now we are free to re-regularize the integral (\ref{eq:I2}) at $\,d=4\,$
 by placing a common physical momentum cutoff $\,\Lambda$\,,
 \begin{equation}
 {\cal I}_{2} ~=\, \int^{\Lambda}\!\frac{d^{4}k}{(2\pi)^4}\frac{1}{k^{2}}
 ~=\, -i\frac{\Lambda^{2}}{16\pi^{2}} \,.
 \end{equation}
 Thus, we can deduce the QED gauge-coupling renormalization,
 \beqa
 \label{eq:1/g2-1/g20}
 \f{1}{g^2(\mu)} &=&
 \f{1}{g^2(\Lambda)} - \f{\,15\kk (\Lambda^2\! -\mu^2)}{128\pi^2 g^2} \,,
 \eeqa
 under the minimal subtraction scheme, and the corresponding renormalization constant,
 \begin{equation}
 \label{eq:Zg}
 Z_{g} ~=~ 1-\frac{\,15\kappa^{2}(\Lambda^{2}\!-\mu^{2})\,}{256\pi^{2}} \,,
 \end{equation}
 where $\,g(\Lambda )=Z_g g(\mu)\,$ and $\mu$ is the renormalization scale.

 From Eq.\,(\ref{eq:1/g2-1/g20}) or Eq.\,(\ref{eq:Zg}), we finally derive the gauge-invariant
 gravitational {\it power-law correction} to the QED $\beta$-function,
\beqa
\label{eq:beta-G}
\Delta\beta(g,\mu) ~=\, -\frac{15}{128\pi^{2}}(\kappa^{2}\mu^{2})g \,,
\eeqa
which is {\it asymptotically free} and gives $\,a_0^{} = -\f{15}{8}\,$ in
Eq.\,(\ref{eq:beta}).  There is no logarithmic graviton correction
to the gauge coupling $\beta$-function
in the absence of cosmological constant\,\cite{gauge-dep2}, since dimensional counting
shows that gravitational logarithmic corrections could only contribute to dimension-6
operators such as $(D_\mu F^{\mu\nu})^2$ rather than the standard dimension-4 gauge
kinetic term (\ref{eq:S-EM}).  With a nonzero cosmological constant $\Lambda_0$ in
(\ref{eq:S-EH}), the graviton-induced logarithmic correction can appear\,\cite{Toms-CC}
since $\Lambda_0$ has mass-dimension equal 2 and thus the product $\,\Lambda_0\kappa^2\,$
provides a proper dimensionless parameter for one-loop gravitational logarithmic correction.
We stress that due to the nonzero new result in the above Eq.\,(\ref{eq:beta-G}),
{\it the graviton-induced leading power-law correction will eventually
dominate gauge coupling running at high scales and always drive gauge unification nearby the
Planck scale, irrespective of the detail of all logarithmic corrections.}
This will be demonstrated in the next section.

 For comparison, we want to clarify how our above analysis differs from that of the
 conventional BFM (or the equivalent diagrammatical) approach.
 The latter corresponds to setting all the
 $\Gamma_{jk}^i$ related connection terms in Sec.\,2-3 vanish. In consequence,
 only the diagrams (a)-(b) in Fig.\,\ref{photon-self-energy} survive.
 Then, if we apply the usual naive momentum-cutoff procedure for quadratical divergence,
 we find that Fig.\,\ref{photon-self-energy}(a) and (b) exactly cancel with each other,
 \beqa
 && \mathrm{(a)} ~=\, -\mathrm{(b)}
 ~=~ \frac{\,3(1\!+\!\zeta)\,}{2}\kk
 (p^{2}\eta_{\mu\nu}^{}\! -p_{\mu}^{}p_{\nu}^{})\,{\cal I}_2
 \,,~~~~~~~~~~
\nn\\[1.5mm]
 && \mathrm{(a)} + \mathrm{(b)} ~=~ 0 \,.
 \eeqa
 This also agrees to the null result of the conventional diagrammatic calculation in the
 second paper of \cite{gauge-dep2}. As another check, we can apply DRED method to regularize
 Fig.\,\ref{photon-self-energy}(a)-(b) by setting all connection terms vanish. Then we
 find the two diagrams no longer cancel, but their sum depends on graviton gauge-parameter
 $\,\zeta$\, and thus non-physical,
 \beqa
 &&
 \mathrm{(a)} ~=~ 3\zeta \kk (p^{2}\eta_{\mu\nu}^{}\! -p_{\mu}^{}p_{\nu}^{})
 \,{\cal I}_2\,,
 \nn\\[1.5mm]
 &&
 \mathrm{(b)} ~=\, -(3+\zeta ) \kk
 (p^2\eta_{\mu\nu}^{}\! -p_{\mu}^{}p_{\nu}^{}) \,{\cal I}_2 \,,
 \nn\\[1.5mm]
 &&
 \mathrm{(a)} + \mathrm{(b)} ~=~ (-3+2\zeta ) \kk
 (p^{2}\eta_{\mu\nu}^{}\! -p_{\mu}^{}p_{\nu}^{})\,{\cal I}_2 \,.
 \hspace*{15mm}
 \eeqa
 This shows that even for well-defined gauge-invariant DRED regularization for quadratical
 divergence (\'{a} la Veltman\,\cite{Veltman}), it is crucial to further use
 the Vilkovisky-DeWitt effective action
 (including the connection-induced new contributions to Fig.\,\ref{photon-self-energy}(a)-(e))
 for ensuring the {\it full gauge-invariance.}
 This also explains why the gauge-invariant nonzero power-law correction (\ref{eq:beta-G})
 was not discovered before.

 As one more consistency check of the present calculation, we have used
 another way\,\cite{Toms-CC} to compute the Vilkovisky-DeWitt effective action
 in the coordinate space and with the aid of Wick theorem.
 Let us define the photon and graviton propagators,
\begin{subequations}
\beqa
\langle a_\mu^{}(x)a_\nu^{}(y)\rangle &\!=\!& D_{\mu\nu}(x,y) \,,
\\[1.5mm]
\langle h_{\mu\nu}(x)h_{\alpha\beta}(y)\rangle
 &\!=\!& D_{\mu\nu,\alpha\beta}(x,y) \,,
\eeqa
\end{subequations}
with
\begin{subequations}
\beqa
D_{\mu\nu}(x,y) &\!=\!& \int\!\!\frac{d^4k}{(2\pi)^4}e^{-ik(x-y)}D_{\mu\nu}(k) \,,
\\
D_{\mu\nu,\alpha\beta}(x,y)
&\!=\!& \int\!\!\frac{d^4k}{(2\pi)^4}e^{-ik(x-y)}D_{\mu\nu,\alpha\beta}(k) \,.~~~~~~
\eeqa
\end{subequations}
Thus, we derive the effective action for gauge field,
\begin{equation}
i\Gamma_{\rm A} ~=~ \langle iS_2\rangle-\frac{1}{2}\langle S^2_1\rangle
\end{equation}
 where $S_{1}$ and $S_{2}$ are the action terms containing
 one and two external fields, respectively.
 Then we compute the effective action for Landau-DeWitt gauge by using the
 {\sc Cadabra} package\,\cite{Cadabra}, and deduce the gauge-part,
 \begin{equation}
 i\Gamma_{\rm A} ~=
 -\frac{\,7\kk\,}{8}{\cal I}_2
 \!\int\!\!d^4x\,\f{1}{4}\bar{F}_{\mu\nu}\bar{F}^{\mu\nu} \,,
 \end{equation}
 corresponding to diagrams in Fig.\,\ref{photon-self-energy}(a)-(c), and
 the sum of Eqs.\,(\ref{eq:Ph-SE-a})-(\ref{eq:Ph-SE-c}).
 For ghost-part, we derive
 \beqa
 i\Gamma_{\rm gh} &\!=\!&
 \langle iS_{2\rm gh}\rangle - \f{1}{2}\left<S_{1\rm gh}^2\right>
\nn\\[1mm]
 &\!=\!&   -\kappa^2{\cal I}_{2}
 \!\int\!\!d^4x\, \f{1}{4}\bar{F}_{\mu\nu}\bar{F}^{\mu\nu} \,,~~~~~~~~
 \eeqa
 which corresponds to diagrams in Fig.\,\ref{photon-self-energy}(d)-(e), and
 the sum of Eqs.\,(\ref{eq:Ph-SE-d})-(\ref{eq:Ph-SE-e}).
 With these we obtain the full one-loop effective action,
 \beqa
 \label{eq:Gamma-Wick}
 \Gamma ~=~ \Gamma_{\rm A} + \Gamma_{\rm gh} ~=\,
 -\frac{\,15\kappa^2\Lambda^2\,}{\,128\pi^2\,}
  \!\int\!\!d^4x\, \f{-1\,}{4}\bar{F}_{\mu\nu}\bar{F}^{\mu\nu} \,,~~~~~
 \eeqa
 from which we reproduce the same gauge coupling renormalization as in
 (\ref{eq:1/g2-1/g20})-(\ref{eq:Zg}) and
 the same power-law correction to the $\beta$ function as in (\ref{eq:beta-G}).

\vspace*{3mm}

 Next, we discuss the extension of the above analysis to
 non-Abelian gauge theories coupled to Einstein gravity.
 We first note that a non-Abelian gauge theory adds no more graviton-induced
 one-loop self-energy diagram beyond those given in Fig.\,\ref{photon-self-energy}, except the
 couplings in these diagrams may differ from QED. So, let us inspect the possible change
 in each diagram of Fig.\,\ref{photon-self-energy} for the non-Abelian case.

 First, Figs.\,\ref{photon-self-energy}(a) and \ref{photon-self-energy}(d) contain
 pure gravitational interactions only, so they remain the same for non-Abelian theories.
 Second, Figs.\,\ref{photon-self-energy}(b) and \ref{photon-self-energy}(e) do not change
 too, since both graviton and graviton-ghost carry no gauge-charge. So both the gauge fields
 in and outside the loop must share the same ``color'' and thus no extra summation of ``color"
 over the loop gauge-field.
 Third, Fig.\,\ref{photon-self-energy}(c) could receive a change due to possible ``color"
 summation over the gauge-loop.
 The relevant changes would come from two places at one-loop level.
 One is the gauge-fixing (\ref{gauge-condition-b}),
 which contributes a term of the following form,
 \beqa
 \bar F^a_{\mu\nu}\bar F^{b}_{\rho\sigma} a^{a}_\alpha a^{b}_\beta \,.
 \eeqa
 Thus, given the two external background gauge-fields (from
 $\,\bar{F}^a_{\mu\nu}\bar{F}^{b}_{\rho\sigma}\,$) for Fig.\,\ref{photon-self-energy}(c), there
 is no more summation over the gauge-indices of the fluctuating gauge-field in the loop.
 The other contribution comes from the connection term, namely the
 first term in (\ref{eq:L-con}),
 which would contribute to Fig.\,\ref{photon-self-energy}(c) via the form,
 \beqa
 \bar F^a_{\mu\nu}\bar F^{a}_{\rho\sigma} a^{b}_\lambda a^{b}_\kappa \,.
 \eeqa
 This allows a summation over the loop gauge-indices ``$b$" and
 enhances the contribution by an overall factor of the number
 of non-Abelian gauge fields,  which equals $\,N^2-1\,$ for the $SU(N)$ gauge group.
 But our explicit calculation shows that this connection-induced contribution
 actually vanishes for both Abelian and non-Abelian cases.
 Hence, the conclusion is that our graviton induced power-law correction (\ref{eq:beta-G})
 is {\it universal for both Abelian and non-Abelian gauge theories.}
 This means that the same coefficient $\,a_0^{}\,$ in (\ref{eq:beta})
 holds for all gauge couplings.

 \section{4. Gravity Assisted Gauge Unification}

 Gauge coupling unification is a beautiful idea that suggests the three apparently different gauge couplings
 of the SM (as measured at low energies) would converge to a single coupling of the
 grand unification (GUT) group\,\cite{SU5} at high scales. The evolutions of gauge couplings from low scale
 to GUT scale are conventionally governed by the renormalization group eqtaions (RGEs)
 with logarithmic running\,\cite{GGUT}.
 The precision data show that logarithmic evolutions of the three gauge couplings do not exactly converge
 for the SM particle spectrum\,\cite{PDG}, while the convergence works fine in
 the minimal supersymmetric standard model (MSSM) with one-loop RG running.
 But more precise numerical analyses
 including two-loop RG running reveal that even in MSSM the strong gauge coupling $\,\alpha_3^{}\,$
 does not exactly meet with the other two at the GUT scale as its value is smaller than
 $(\alpha_1^{},\,\alpha_2^{})$ by $\sim\!3\%$ (a $5\sigma$ deviation);\,
 so it is necessary to carefully invoke the model-dependent
 one-loop threshold effects\,\cite{GUTrev}.
 It was also argued that the gravity-induced effective higher
 dimensional operators can generate uncertainties larger than the usual two-loop effects of MSSM
 and thus significantly alter the gauge unification\,\cite{Calmet}.

\begin{figure}[h]
   \begin{center}
    {\includegraphics[width=0.48\textwidth]{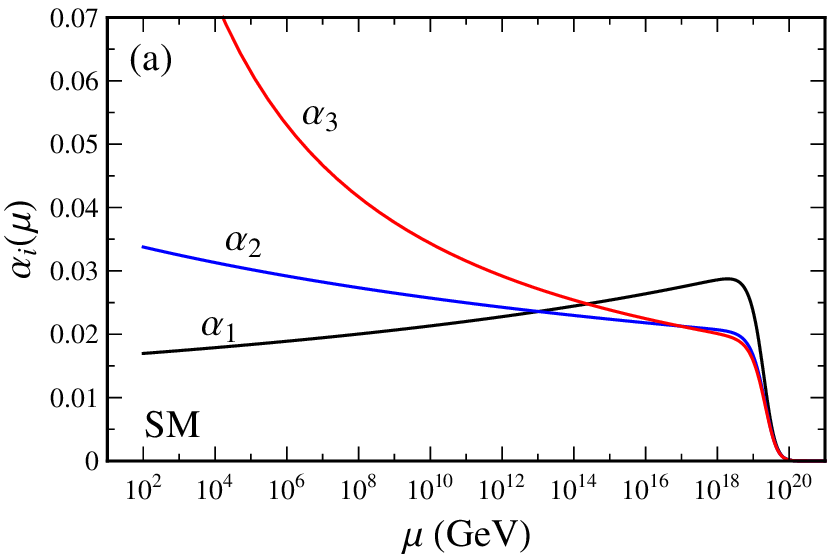}}
    {\includegraphics[width=0.48\textwidth]{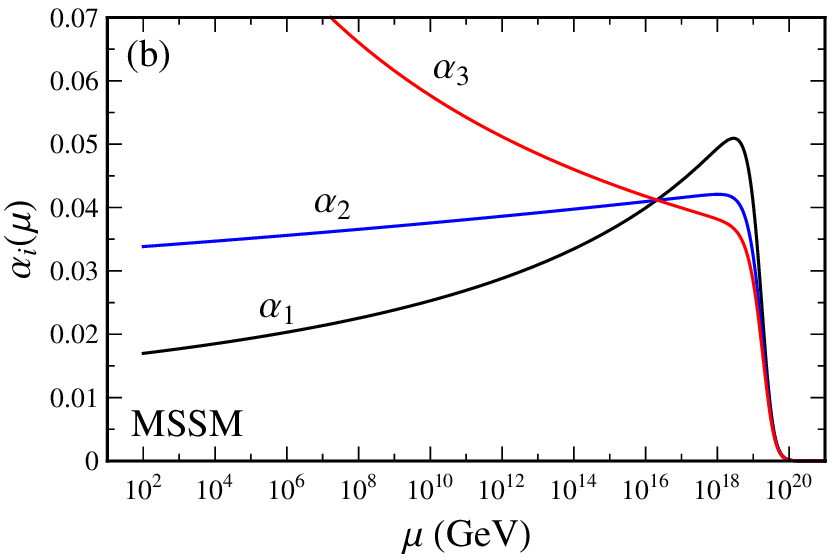}}
   \end{center}
   \vspace*{-5mm}
   \caption{Gravity assisted gauge unifications in the SM [plot-(a)] and in the MSSM [plot-(b)].
   In each case the graviton-induced universal power-law runnings become dominant above $10^{18}$\,GeV
   and always drive three gauge couplings to rapidly converge towards the UV fixed point,
   nearby the Planck scale.}
   \label{fig:GUnify}
 \end{figure}

 With the gauge-invariant gravitational power-law corrections (\ref{eq:beta-G}), we can resolve
 running gauge coupling $\,\al_i(\mu )=g_i^2(\mu)/4\pi\,$ from the RGE (\ref{eq:beta}),
\beqa
\label{eq:RG-Sol-alpha-i}
\f{\,e^{-c\mu^2}}{\al_i^{}(\mu)} - \f{\,e^{-c\mu^2_0}}{\al_i^{}(\mu_0^{})} ~=~
\f{b_{0i}}{4\pi}\int_{\mu_0^2}^{\mu^2}\! \f{dx}{x}e^{-cx} \,,
\eeqa
 with $\dis\,c\equiv\f{|a_0|\kk}{(4\pi)^2}=\f{15}{8\pi M_P^2}\,$.\,
 So we further deduce,
 \beqs
 \label{eq:alpha-i}
 \beqa
 \label{eq:alpha-i1}
 \al_i^{}(\mu ) &\!\!=\!\!& \f{\al_i^{}(\mu_0^{})e^{-c\mu^2}}{\,e^{-c\mu_0^2}
                +\f{b_{0i}\al_{i}(\mu_0^{})}{4\pi}\int_{\mu_0^2}^{\mu^2}\! \f{dx}{x}e^{-cx}\,}
 \\[3mm]
&\!\!\simeq\!\!& \f{\al_i^{}(\mu_0^{})\,e^{-c\mu_0^2}}{\,1
                +\f{b_{0i}\al_i^{}(\mu_0^{})}{4\pi}
                \[\ln\f{\mu^2}{\mu_0^2} - c\mu^2\]\,} \,,
\label{eq:alpha-i2}
 \eeqa
 \eeqs
where for the estimate in (\ref{eq:alpha-i2}) we have kept in mind that $\,\mu_0^{}\lll M_P\,$
(such as the choice of $\,\mu_0^{}=M_Z\,$ below), and we also expanded the exponential integral
for $\,\mu < c^{-\f{1}{2}}\simeq 1.3M_P\simeq 2\times 10^{19}\,$GeV
(which holds for most energy regions in Fig.\,\ref{fig:GUnify}).\,
Eq.\,(\ref{eq:alpha-i}) explicitly shows that the evolution of any gauge coupling $\,\al_i^{}(\mu)\,$
will be exponentially suppressed by
$\,\exp\!\[-c\mu^2\]\,$, which dominates the running behavior for high scales
above $\,O(10^{-2}M_P^2)$\,.\,
Hence, the universal gravitational power-law corrections will
{\it always drive all gauge couplings to rapidly converge to the UV fixed point
at high scales and reach unification around the Planck scale,
irrespective of the detail of their logarithmic corrections and initial values.}
This feature is numerically demonstrated in Fig.\,\ref{fig:GUnify}
by using the evolution equation (\ref{eq:alpha-i1}).

In Fig.\,\ref{fig:GUnify}(a)-(b), we have analyzed the gauge coupling runnings
for both the SM and the MSSM,
where the conventional coefficients $b_{0i}$ in the one-loop RGEs are,
$\,(b_{01},\,b_{02},\,b_{03})=\(-\f{41}{10},\,\f{19}{6},\,7\)\,$ for the SM and
$\,(b_{01},\,b_{02},\,b_{03})=\(-\f{33}{5},\,-1,\,3\)\,$  for the MSSM.
(Here we have adopted the conventional MSSM particle spectrum without including
the superpartner of graviton -- the spin-$\f{3}{2}$ gravitino, whose potential contribution
to the gauge coupling running is worth of a future study.)
We have also input the initial values,
$\,\alpha_1^{-1}=59.00\pm 0.02,\,$
$\,\alpha_2^{-1}=29.57\pm 0.02,\,$ and
$\,\alpha_3^{-1}=8.50\pm 0.14\,$, at the $Z$-pole $\,\mu_0^{}=M_Z\,$ \cite{PDG}.

For the graviton-induced one-loop $\beta$-function (\ref{eq:beta}) with $\,a_0^{}\,$ given in
(\ref{eq:beta-G}), we note that at the Planck scale $\,\mu =M_P\,$,\,
the loop-expansion parameter,
$\,\f{|a_0^{}|}{(4\pi)^2}(\kk\mu^2) = \f{15}{8\pi}\f{\mu^2}{M_P^2}
 \simeq 0.6 < 1\,$,\, and the condition
$\,\f{|a_0^{}|}{(4\pi)^2}(\kk\mu^2)< 1\,$ holds for
$\,\mu < 2\times 10^{19}\,$GeV,  which is better than the naive expectation.
Fig.\,\ref{fig:GUnify} shows that above $10^{19}$\,GeV the gauge couplings rapidly converge
due to the exponential suppression in (\ref{eq:alpha-i1});
and quite before approaching the UV fixed point
the three curves become indistinguishable for $\,\al_i^{}\lesssim 0.01\,$ (SM)
and $\,\al_i^{}\lesssim 0.02\,$ (MSSM), corresponding to a scale of about
$\,2\times 10^{19}$\,GeV.  Given the experimental error of
$\,\al_3(M_Z)=(8.50\pm 0.14)^{-1}$\, and potential higher loop effects,
this is enough for a possible consistent unification with a finite value of coupling
and at a single scale around the Planck energy.
For the perturbative one-loop running of gauge couplings, the renormalization of
$\kk$ belongs to higher order effect here and is thus not included.
It is useful to note that there are clear evidences supporting
Einstein gravity to be asymptotically safe via the existence of nontrivial UV fixed point in its RG flow,
so it may be UV complete and perturbatively sensible even beyond the Planck scale\,\cite{Asafe1,Asafe2}.
To be conservative, we would consider the one-loop prediction of Fig.\,\ref{fig:GUnify} as an
instructive extrapolation of the Einstein gravity, since this provides an encouraging insight
on the important role of gravity for realizing the gauge unification, and should strongly
inspire and motivate more elaborate investigations along this direction.

With these said, we make a few more comments.
Fig.\,\ref{fig:GUnify}(a) shows that for the SM gauge coupling evolutions,
despite their familiar non-convergence in the region of $10^{13-17}$\,GeV, the three couplings do unify
around the Planck scale.
Mathematically, as shown in Eq.\,(\ref{eq:alpha-i}),
to make $\al_i(\mu)$ reach the UV fixed point would
require $\,\mu\to \infinity\,$;  but practically, due to the one-loop exponential suppression in
(\ref{eq:alpha-i}), the differences among $\al_i(\mu)$'s will rapidly decrease to be smaller than the
experimental error of $\,\al_3(M_Z)=(8.50\pm 0.14)^{-1}\,$,
this is enough for a possible consistent unification with a finite value of coupling
and at a single scale around the Planck energy.
Also, taking into account of higher loop corrections and possible threshold effects
may well fill in any remaining tiny gap among $\al_i(\mu)$'s at this scale.
For the MSSM, as shown in Fig.\,\ref{fig:GUnify}(b),
one needs not to worry about the model-dependent threshold effects
or the two-loop-induced non-convergence around the scale of $10^{16}$\,GeV, it is quite possible
that the GUT does not happen around the scale of $10^{16}$\,GeV, as in the SM case. Instead,
the real GUT would be naturally realized around the Planck scale,
and thus is expected to simultaneously unify
with the gravity force as well. This also removes the old puzzle on why the conventional GUT scale
is about three orders of magnitude lower than the fundamental Planck scale.
Furthermore, the Planck scale unification helps to sufficiently postpone nucleon decays,
which explains why all the experimental data so far support the proton stability.
In addition, this is also a good news for various approaches of dynamical electroweak symmetry
breaking\,\cite{TC}, such as the technicolor type of theories,
which often invoke many gauge groups with new strong forces at the intermediate scales and
makes one worry about whether a gauge unification could ever be realized at certain
high scales in these theories.
Fortunately, the universal gravitational power-law running for all gauge couplings
found above should drive a final unification at the Planck scale.
It suggests that the {\it Planck scale unification} may be
a generic feature of all low energy gauge groups
and is fully consistent with the experimental evidences of proton stability.
It is also an appearing feature of having
the possibility of a simultaneous unification
of all four fundamental interactions around the Planck scale.

 \section{5. Gravitational Correction to Higgs Boson Coupling and Mass}

\begin{figure}[h]
  \begin{center}
  \includegraphics[width=0.44\textwidth]{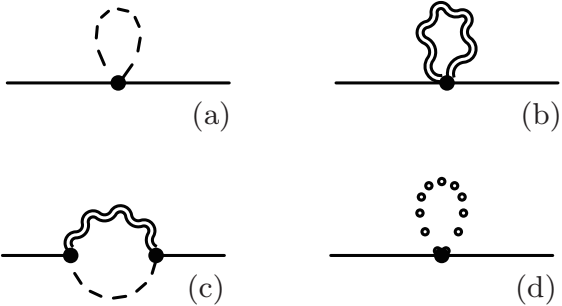}
    %
    \vspace*{-2mm}
    \caption{Graviton-induced self-energy for scalar field: dashed-line is fluctuating scalar field,
    double-wavy-line denotes graviton, and circle-dotted-line depicts graviton-ghost.}
    \label{fig:scalar2}
  \end{center}
 \end{figure}

\begin{figure}[h]
  \begin{center}
  \includegraphics[width=0.48\textwidth]{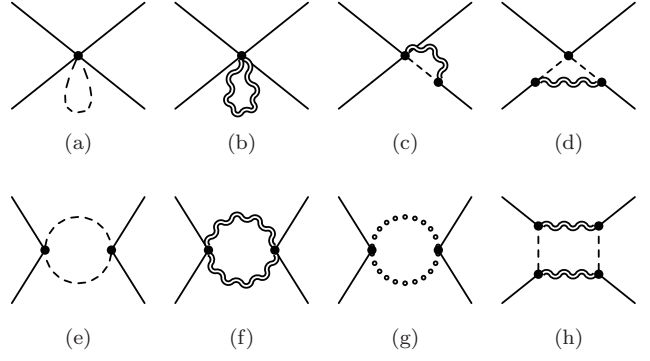}
    \caption{Graviton-induced corrections to quartic scalar vertex:
    dashed-line is fluctuating scalar field,
    double-wavy-line denotes graviton, and circle-dotted-line depicts graviton-ghost.}
    \label{fig:scalar4}
  \end{center}
 \end{figure}

 Without losing generality,
 we first consider a real scalar field which minimally couples to
 Einstein gravity,
\beqa
\label{eq:S-phi4}
 S_\phi ~= \int\!\!d^4x\,
\sqrt{-g}\[\frac{1}{2}g^{\mu\nu}\partial_\mu\phi\partial_\nu\phi-V(\phi)\]\!,~~~~
\eeqa
 where $V(\phi)$ is the scalar potential,
 $\, V = \f{1}{2}m^{2}\phi^{2}+\frac{\lambda}{4!}\phi^{4}$,\,
 with $\,m^2\geqq 0\,$ ($\,m^2< 0\,$) corresponding to the unbroken (broken) phase.
 To compute the effective potential of scalar field,
 we expand the graviton and scalar fields around their backgrounds,
\begin{equation}
g_{\mu\nu}^{} ~=~ \eta_{\mu\nu}^{} + \kappa h_{\mu\nu}^{} \,,~~~~~
\phi~=~\bar{\phi}+\phihh \,.
\end{equation}
 We impose the Landau-DeWitt gauge condition for the gauge-fixing,
 \beqa
 {\cal L}_{\rm gf} ~=~
 \f{1}{2\zeta}\(h^{\mu}_{\nu,\mu} - \f{1}{2}h_{,\nu}
 - \f{\k}{2}\phihh\,\partial_\nu^{}\phibar\)^2 .
 \eeqa
 Then, we derive the connection-induced terms in the graviton-scalar sector
 of the Lagrangian,
 \beqa
 \label{eq:Lconn-S}
 {\cal L}_{\rm conn} \,&=&\,
 -\f{\kk}{2}S_{,g_{\mu\nu}^{}}\Gamma^{g_{\mu\nu}^{}}_{g_{\alpha\beta}g_{\rho\sigma}}
            h_{\alpha\beta}h_{\rho\sigma}
 -\f{1}{2}S_{,g_{\mu\nu}^{}}\Gamma^{g_{\mu\nu}^{}}_{\phi\phi}\phihh^2\nonumber\\
 & &- \k S_{,\phi}\Gamma^{\phi}_{\phi g_{\mu\nu}^{}}\phihh \;h_{\mu\nu}^{}
 \,,~~~~
 \eeqa
 with
 \beqs
 \beqa
 S_{,g_{\mu\nu}^{}} &\!\!=\!&
 -\f{1}{2}\dif^\mu\phibar\dif^\nu\phibar +\f{1}{2}\eta^{\mu\nu}
 \[ \f{1}{2}\dif_\rho\phibar\dif^\rho\phibar  - V(\phibar ) \] \!,~~~~~~~~~
 \\[1mm]
 S_{,\phi} &\!\!=\!&
 -\dif^2\phibar - m^2\phibar -\f{\lambda}{6}\phibar^3 \,.
 \eeqa
 \eeqs
 In (\ref{eq:Lconn-S}) the graviton-field connection
 $\,\Gamma^{g_{\mu\nu}^{}}_{g_{\alpha\beta}g_{\rho\sigma}}\,$
 was given in Eq.\,(\ref{connection-qed}), and the scalar-field related connections are,
 \beqs
 \beqa
 \Gamma_{\phi(y)\phi(z)}^{g_{\mu\nu}^{}(x)}
 &\!\!=\!& \f{\,\kk}{4}g_{\mu\nu}^{}\delta (x\!-\!y)\delta (x\!-\!z) \,,
 \hspace*{15mm}
 \\[1mm]
 \Gamma_{\phi(y)g_{\mu\nu}^{}(z)}^{\phi(x)}
 &\!\!=\!& \f{1}{4}g^{\mu\nu}\delta (x\!-\!y)\delta (x\!-\!z) \,,
 \\[1mm]
 \Gamma_{\phi(y)\phi(z)}^{\phi(x)} &\!\!=\!&
 \Gamma_{\phi(y)g_{}^{\mu\nu}(z)}^{g_{\al\be}(x)} \,=~ 0 \,,
 \eeqa
 \eeqs
 which are derived from the metrics
 $\,G_{\phi (x)\phi (y)} = \sqrt{-g}\,\delta (x\!-\!y)\,$ and
 $\,G_{g_{\mu\nu}^{}(x)g_{\al\be}^{}(y)}\,$ [as in Eq.\,(\ref{gravity-metric})].
 The Feynman diagrams for the graviton-induced
 scalar self-energy corrections and quartic vertex corrections
 are shown in Fig.\,\ref{fig:scalar2} and Fig.\,\ref{fig:scalar4}, respectively.


 For the scalar self-energy corrections,
 we compute the quadratic divergent part for each diagram in Fig.\,\ref{fig:scalar2},
 \begin{subequations}
 \label{eq:phi-SF}
 \beqa
 \label{eq:phi-SF-a}
 \mathrm{Fig.4(a)} &\!=\!&
 \[ \f{1}{2}\lambda+\kappa^{2}\(\f{1}{8}-\frac{1}{4\zeta}\)p^{2}
   -\f{1}{4}\kappa^{2}m^{2}\]\II_{2} \,,~~~~~~~~~
 \\[1.5mm]
 \label{eq:phi-SF-b}
 \mathrm{Fig.4(b)} &\!=\!& -3\kappa^{2}m^{2}\;\II_{2} \,,
 \\[1.5mm]
 \label{eq:phi-SF-c}
 \mathrm{Fig.4(c)} &\!=\!& \(-1+\frac{1}{4\zeta}\)\kappa^{2}p^{2}\;\II_{2} \,,
 \\[1.5mm]
 \label{eq:phi-SF-d}
 \mathrm{Fig.4(d)} &\!=\!& \kappa^{2}p^{2}\;\II_{2} \,,
 \eeqa
 \end{subequations}
 where the $\lambda$-term in (\ref{eq:phi-SF-a}) comes from the scalar quartic
 self-interaction alone, which we have included for the comparison with graviton
 induced corrections and for the convenience of analysis.
 We also note that unlike the case of photon self-energy, the external lines of
 the scalar self-energy carry no Lorentz index, so the corresponding loop integrals
 can be directly reduced to a scalar integral without $\,k^\mu k^\nu\,$ like combination
 of loop-momenta, and thus no symmetrization of loop-momenta is needed.

 For the graviton-induced vertex corrections in Fig.\,\ref{fig:scalar4},
 only the first two diagrams have quadratical divergence,
 \begin{subequations}
 \beqa
 \mathrm{Fig.5(a)} &\!=\!& -\frac{1}{4}\lambda\kappa^{2}\II_{2} \,,
 \\[1mm]
 \mathrm{Fig.5(b)} &\!=\!& -3\lambda\kappa^{2}\II_{2} \,,
 \eeqa
 \end{subequations}
 and all other diagrams contain logarithmic divergence at most.

 Summing up relevant contributions we deduce the two-point proper self-energy
 and four-point proper vertex for scalar field,
 \beqs
 \label{eq:phi-2point4point}
 \beqa
 \label{eq:phi-2point}
 \Gamma_2(p) &\!=\!&
 \[\f{1}{2}\lambda + \f{1}{8}(p^2 - 26 m^{2})\kk\]\!\II_{2} \,,
 \\[1mm]
 \label{eq:phi-4point}
 \Gamma_4(p) &\!=\!&
 -i\lambda-\f{13}{4}\lambda\kappa^{2}\;\II_{2} \,,
 \eeqa
 \eeqs
 where we do not include the graviton-induced logarithmic divergent terms
 (as in \cite{Toms-phi}) because they are negligible as compared to the
 dominant power-law corrections. Note that the sum of
 (\ref{eq:phi-SF-a})-(\ref{eq:phi-SF-d}) explicitly proves the exact cancellation
 of the $\,\f{1}{\zeta}\,$ gauge-parameter poles, which is a consistency check of
 our Landau-DeWitt gauge calculation.

 To fully understand the results (\ref{eq:phi-2point})-(\ref{eq:phi-4point}), we have
 compared them with those in the conventional BFM (or the equivalent diagrammatical)
 approach where all connection-induced new terms are set to zero.
 The findings are summarized in Table-1.

 From Eqs.\,(\ref{eq:phi-2point})-(\ref{eq:phi-4point}),
 we derive the renormalization for scalar coupling,
 $\,\lambda (\mu) = Z_\phi^2 Z_\lambda^{-1}\lambda(\Lambda)\,$,\,
 with the following renormalization constants,
 \begin{subequations}
 \label{eq:Zphi-Zlambda}
 \beqa
 \label{eq:Zphi}
 Z_{\phi} &\!=\!&
 1+\frac{1}{128\pi^{2}}\kappa^{2}(\Lambda^{2}-\mu^{2}) \,,
 \\[1mm]
 \label{eq:Zlambda}
 Z_{\lambda} &\!=\!&
 1+\frac{13}{64\pi^{2}}\kappa^{2}(\Lambda^{2}-\mu^{2}) \,,
 \eeqa
 \end{subequations}
 under the minimal subtraction scheme.
 Then, with (\ref{eq:Zphi-Zlambda}) we compute the graviton-induced scalar $\beta$-function,
 \beqa
 \label{eq:beta-phi}
 \Delta\beta(\lambda,\mu) ~=\, 
 +\frac{3\lambda}{8\pi^{2}}(\kappa^{2}\mu^{2}) \,,
 \eeqa
 which is {\it not} asymptotically free, contrary to the gauge coupling $\beta$-function
 (\ref{eq:beta-G}) we derived earlier. The pure scalar loop correction is logarithmically
 divergent and its renormalization gives the usual non-asymptotically free scalar
 $\beta$-function  $\,\beta_0 = +\f{3\lambda^2}{16\pi^2}\,$.

 \begin{widetext}
 \begin{table*} 
 \renewcommand{\arraystretch}{1.25}
 \tabcolsep 6pt
 \begin{center}
\begin{tabular}{c||c|c|c}
\hline\hline
Graviton-Induced
 & Conventional & Connection-Induced & Sum (Our Results)
 \\
Corrections  & Approach     & only (from VDA)  & (Landau-DeWitt $\zeta\to 0$)
 \\
\hline\hline
Fig.4(a)
& $0$ & $(\frac{1}{8}-\frac{1}{4\zeta})p^{2}-\frac{1}{4}m^{2}$
& $(\frac{1}{8}-\frac{1}{4\zeta})p^{2}-\frac{1}{4}m^{2}$\\
\hline
Fig.4(b)
& $-(3+2\zeta)m^{2}$ & $0$ & $-3m^{2}$\\
\hline
Fig.4(c)
& $\zeta p^{2}$ & $(-1+\frac{1}{4\zeta})p^{2}$ & $(-1+\frac{1}{4\zeta})p^{2}$\\
\hline
Fig.4(d)
& $0$ & $p^{2}$ & $p^{2}$\\
\hline
$\Gamma_2$ & $\zeta p^{2}-(3+2\zeta)m^{2}$ & $\frac{1}{8}p^{2}-\frac{1}{4}m^{2}$
& $\frac{1}{8}(p^{2}- 26m^{2})$
\\
\hline\hline
Fig.5(a) & $0$ & $-\frac{1}{4}\lambda$ & $-\frac{1}{4}\lambda$
\\
\hline
Fig.5(b) & $-(3+2\zeta)\lambda$ & $0$ & $-3\lambda$
\\
\hline
$\Gamma_4$ & $-(3+2\zeta)\lambda$ & $-\frac{1}{4}\lambda$ & $-\frac{13}{4}\lambda$
\\
\hline\hline
\end{tabular}
\end{center}
\vspace*{-4mm}
 \caption{Graviton-induced power-law corrections to scalar self-energy and quartic vertex:
 Summary of the comparison between the conventional approach
 (with vanishing connection and with general gauge-parameter $\zeta$)
 and Vilkovisky-DeWitt approach (VDA)
 (with nonzero connection, and in Landau-DeWitt gauge
 with $\,\zeta\to 0\,$ in the end). In each entry of the contribution, a common factor
 $\,\kappa^2\II_2\,$ is factorized out. Our summed results of $\Gamma_2$ and $\Gamma_4$
 on the 3rd column agree with Eqs.\,(\ref{eq:phi-2point})-(\ref{eq:phi-4point}).}
 \end{table*}
 \end{widetext}

 From the two-point proper self-energy (\ref{eq:phi-2point}), we further perform the
 renormalization for scalar mass in the on-shell scheme, which fixes the mass counter term,
 \beqa
 \label{eq:phi-mass-CT}
 \delta m^2 ~=~
 \frac{1}{32\pi^{2}}\(-\lambda +\frac{25}{4}m^{2}\kappa^{2}\)
 \Lambda^{2} \,,
 \eeqa
 where we have defined the renormalized mass, $\,m^2 = m_0^2 - \delta m^2 \,$
 with $m_0^{}$ denoting the bare mass parameter in the original Lagrangian,
 and also included the contribution from the pure scalar loop.
 Comparing the two terms on the right-hand-side of (\ref{eq:phi-mass-CT}), we note
 that the graviton-induced quadratical divergence is actually much softer since
 the product $\,\kk\cut^2=16\pi(\cut /M_P)^2 = O(10^2)\,$ for an ultraviolet cutoff
 $\,\cut\sim M_P\,$.\,

 It is straightforward to extend the above analysis to the SM Higgs boson,
 since graviton coupling to scalar fields is universal. Let us write down the
 SM Higgs doublet,
 \beqa
 \Phi ~=~ \f{1}{\sqrt{2}\,}
 \(\!\ba{c} \pi_1^{} + i\pi_2^{} \\[1mm] \sigma +i\pi_3^{} \ea\!\) ,
 \eeqa
 where $\,\dis\Phi^\dag\Phi = \si^2 + \!\sum_{a=1}^3\pi_a^2\,$
 and $\,\sigma = \bar{\sigma}+\hat{\sigma}$\,,\, with
 $\,\sigmah\,$ being the SM Higgs boson and $\,\pi_{1,2,3}^{}\,$
 the would-be Goldstone bosons.
 The Higgs background field $\,\sigmab\,$  will equal the vacuum expectation value (VEV),
 $\,v\,$, at the minimum of Higgs potential.

 Consider the SM Higgs potential, $\,V=m^2(\Phi^\dag\Phi)+\lambda (\Phi^\dag\Phi)^2\,$,\,
 with $\,m^2<0\,$ and the VEV, $\,v =\sqrt{-m^2/\lambda}\,$.\,
 Then we recompute the gravitational power-law corrections to Higgs boson
 self-energy and quartic vertex in Fig.\,\ref{fig:scalar2}-\ref{fig:scalar4},
 and deduce the following,
 \beqs
 \label{eq:Higgs-2point4point}
 \beqa
 \label{eq:Higgs-2point}
 \Gamma_{H2}^{}(p) &\!=\!&
 \[\f{1}{2}\lambda + \hf{(p^2 - 8 m^2_{H})\kk}\]\!\II_{2} \,,~~~~~~~~
 \\[1mm]
 \label{eq:Higgs-4point}
 \Gamma_{H4}^{}(p) &\!=\!&
 -i\lambda - 24\lambda\kappa^{2}\;\II_{2} \,.
 \eeqa
 \eeqs
 With these we derive the graviton-induced contributions to the Higgs boson $\beta$-function
 and the Higgs mass counter term,
 \beqs
 \beqa
 \label{eq:beta-H}
 \Delta\beta(\lambda,\mu) &\!=\!&
 +\frac{3\lambda}{8\pi^{2}}(\kappa^{2}\mu^{2}) \,,
 \\[1mm]
 \label{eq:H-mass-CT}
 \delta m^2_H &\!=\!&
 \frac{1}{32\pi^{2}}\(-12\lambda + 7m_{H}^2\kappa^{2}\)
 \Lambda^{2} \,,~~~~~~~~~~~~~
 \eeqa
 \eeqs
 where we have used $\,m^{}_{H}\,$ to denote the physical Higgs boson mass.
 We see that (\ref{eq:beta-H}) happens to be the same as in (\ref{eq:beta-phi})
 while (\ref{eq:H-mass-CT}) has different coefficients from (\ref{eq:phi-mass-CT}).
 As compared to the single real scalar case, the changes in the computation of
 (\ref{eq:beta-H})-(\ref{eq:H-mass-CT}) arise from two sources:
 (i)  the Feynman rule for the quartic Higgs vertex in the SM has
 an extra factor $6$ relative to that from the real scalar potential below Eq.\,(\ref{eq:S-phi4});
 (ii) there are additional Goldstone loops in the SM which contribute to
 Fig.\,\ref{fig:scalar2}(a) and Fig.\,\ref{fig:scalar4}(a).

 \begin{figure}[h]
  \begin{center}
  \vspace*{-25mm}
  \includegraphics[width=0.48\textwidth]{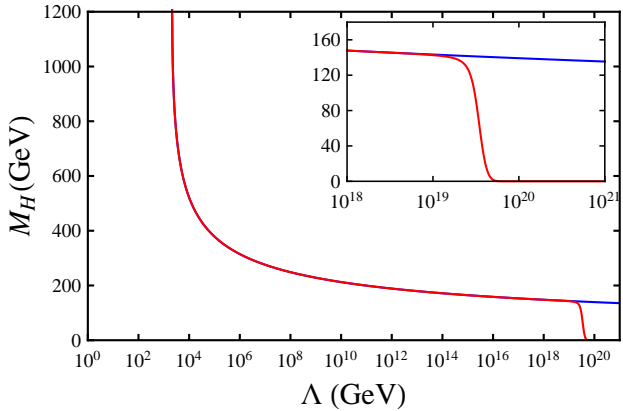}
  \vspace*{-10mm}
    \caption{Gravitational power-law corrections to the triviality bound on the SM Higgs
    boson mass $\,m_H^{}\,$, where the region above each curve is excluded.
    The red curve includes the graviton-induced corrections and
    the blue curve depicts the bound with SM interactions alone.
    }
    \label{fig:TrivialityB}
  \end{center}
 \end{figure}

 Note that the conventional Higgs $\beta$-function in the SM receives only
 logarithmic corrections,
 $\,\beta_0(\lambda,\mu) = +\f{3}{2\pi^2}\lambda^2\,$,\,
 and is not asymptotically free.
 So, with (\ref{eq:beta-H}) we can write the summed Higgs $\beta$ function,
 \beqa
 \label{eq:sum-beta-H}
 \beta(\la,\mu) ~=~ \bt\la^2 + \at\la\kk\mu^2 \,,
 \eeqa
 with $\,(\bt ,\,\at) = \(\f{3}{2\pi^2},\, \f{3}{8\pi^2}\) >0 \,$.\,
 Solving (\ref{eq:sum-beta-H}) we deduce the running Higgs coupling $\,\la(\mu)\,$,
 \beqa
 \label{eq:RGE-Sol-SM}
 \f{e^{\f{\at}{2}\kk\mu^2}}{\lambda(\mu)} ~=~
 \f{e^{\f{\at}{2}\kk\mu^2_0}}{\lambda(\mu_0^{})}
 - \bt \int_{\mu_0^{}}^\mu e^{\f{\at}{2}\kk x^2}\f{dx}{x} \,.
 \eeqa
 We see that when the right-hand-side (RHS) of (\ref{eq:RGE-Sol-SM}) vanishes,
 the renormalized coupling $\lambda(\mu)$ blows up at the Landau pole
 $\,\mu = \cut_L\,$,
 \beqa
 \label{eq:Landau-pole}
 \la^{-1}(\mu_0^{}) ~=~
 \bt e^{-\f{\at}{2}\kk\mu_0^2}\int_{\mu_0^{}}^{\cut_L} e^{\f{\at}{2}\kk x^2}\f{dx}{x} \,.
 \eeqa
 This means that the SM as an effective theory must have an UV cutoff
 $\,\cut < \cut_L\,$.\,
 For a given Higgs boson mass $\,m_H^2 = 2\la (m_H^{})v^2\,$,\,  let us set
 $\,\mu_0^{}=m_H^{}$.\, Thus, from (\ref{eq:Landau-pole}) we can derive the
 triviality bound for Higgs boson mass,
 \beqa
 \label{eq:mH-TB}
 m_H^2 ~<~ \f{2v^2}{\bt} e^{\f{\at}{2}\kk m_H^2}
 \[\int_{m_H^{}}^{\cut}e^{\f{\at}{2}\kk x^2}\f{dx}{x}\]^{-1}\,,
 \eeqa
 where $\,e^{\f{\at}{2}\kk m_H^2} \simeq 1\,$ holds to high accuracy due to the tiny
 factor $\,\f{\at}{2}\kk m_H^2 = \f{3}{\pi}\f{m_H^2}{M_P^2}\simeq 0\,$.\,
 It is clear that in the $\,\at\to 0\,$ limit,
 the condition (\ref{eq:mH-TB}) reduces to the familiar
 triviality bound in the pure SM\,\cite{SM-mH-TB},
 \beqa
 \label{eq:SM-mH-TB}
  m_H^2\ln\f{\cut^2}{m_H^2} ~<~ \f{\,8\pi^2v^2}{3} \,.
 \eeqa
 Keeping this in mind, it is instructive to rewrite our graviton-corrected
 triviality bound (\ref{eq:mH-TB}) as follows,
 \beqs
 \label{eq:mH-TB-2T}
 \beqa
 \label{eq:mH-TB-2}
\hspace*{-12mm}
&&  m_H^2\ln\f{\cut^2}{m_H^2} ~<~ \f{\,8\pi^2v^2}{\,3(1+X)\,} \,,
\\[3mm]
\hspace*{-12mm}
&& X \,\equiv \!\!\int_{m_H^{}}^{\cut}\!\!
   \[e^{\f{\at}{2}\kk x^2}\!-1\]\f{dx}{x}\!\left/ \ln\f{\cut}{m_H^{}} \,>~ 0  \,,~~~~~~
   \right.
\label{eq:mH-TB-X}
 \eeqa
 \eeqs
 where an overall factor
 $\,e^{\f{\at}{2}\kk m_H^2} \simeq 1\,$ on the RHS of (\ref{eq:mH-TB-2}) is safely ignored.
 We find that, because the integral $\,X > 0\,$ generally holds,
 the gravitational power-law corrections always reduce the RHS of (\ref{eq:mH-TB-2}),
 and thus further {\it tighten the triviality bound} relative to (\ref{eq:SM-mH-TB})
 of the pure SM. The graviton-induced corrections play
 a dominant role to enhance the triviality bound
 for the cutoff scale $\,\cut \sim M_P\,$, as clearly shown in Fig.\,6.
 A systematical expansion of the present section (including the power-law corrections to
 Yukawa couplings) will be given in Ref.\,\cite{paper2}.

 \section{6. Conclusions}

The fundamental gravitational force universally couples to all the SM particles and can be
described by the well-defined perturbation expansion
in the modern effective theory formulation\,\cite{EFT-GR}.
The Vilkovisky-DeWitt method\,\cite{Vilkovisky-DeWitt}
profoundly modifies the conventional BFM, and provides the
manifestly gauge-invariant effective action for reliably computing quantum gravity effects.
In this work, we used the Vilkovisky-DeWitt method to derive {\it the first gauge-invariant nonzero
gravitational power-law corrections} to the running of gauge couplings. We found the
gravitational power-law corrections to be universal, {\it making both Abel and non-Abel
gauge couplings asymptotically free} [cf.\ Eq.\,(\ref{eq:beta-G}) and analyses
at the end of Sec.\,3].  We have demonstrated that the graviton-induced
power-law runnings always drive the three SM gauge forces toward to the
UV fixed point, reaching final unification at the Planck scale and irrespective of the detail of
logarithmic corrections (cf. Fig.\,\ref{fig:GUnify}).
This raises the conventional GUT scale by three orders of magnitude, and {\it opens up a natural
possibility of simultaneous unification of all four fundamental gauge forces at the Planck scale.}
We further analyzed the power-law corrections to the $\beta$-function and mass of the SM
Higgs boson [cf.\ Eqs.\,(\ref{eq:beta-H})-(\ref{eq:H-mass-CT})].
We found that the graviton-induced scalar $\beta$-function is not asymptotically
free, and therefore {\it further tightens the triviality bound on the Higgs boson mass,}
as shown in Eq.\,(\ref{eq:mH-TB-2T}) and Fig.\,\ref{fig:TrivialityB}.
Further extensions of the present analysis for computing the power-law corrections
will be given elsewhere\,\cite{paper2}.

\vspace*{5mm}
\noindent
 {\bf Acknowledgments}\\[1mm]
 This research was supported by Chinese NSF (under grants 10625522, 10635030), and
 by the National Basic Research Program of China (under grant 2010CB833000).
 XFW is supported in part by the China Scholarship Council.
 We thank A.\ G.\ Cohen and M.\ E.\ Peskin for stimulating discussions at the
 ``International Conference and Summer School on LHC Physics" (August\,16-25, Beijing).
 After we posted this paper to arXiv:1008.1839, a new paper\,\cite{Toms-new} appeared;
 we thank D.\ J.\ Toms for kindly bringing our attention to his independent study
 via a very different approach (the proper-time method) which reached a similar conclusion
 for graviton corrections to the QED gauge coupling; we also thank him for kind discussion.


\end{document}

Title:
Gauge-Invariant Quantum Gravity Corrections to Gauge Couplings
via Vilkovisky-DeWitt Method and Gravity Assisted Gauge Unification

Authors: Hong-Jian He, Xu-Feng Wang, Zhong-Zhi Xianyu

Abstract:
Gravity is the weakest force in nature, and the gravitational interactions with all
standard model (SM) particles can be well described by perturbative expansions of the
Einstein-Hilbert action as an effective theory, all the way up to energies below the
fundamental Planck scale. We use Vilkovisky-DeWitt method to derive the first gauge-invariant
nonzero gravitational power-law corrections to the running of gauge couplings, which make
both Abel and non-Abel gauge interactions asymptotically free.  We further demonstrate that
the graviton-induced universal power-law runnings always assist the three SM gauge forces
to reach unification around the Planck scale, irrespective of the detail of logarithmic
corrections. We also compute the power-law corrections to the SM Higgs sector and derive
modified triviality bound on the Higgs boson mass.

Comments: 11pp, only minor refinements(fixed some typos, combined eps files for Latex, and
further clarified the use of dimensional reduction method)